# Anisotropic thermal expansion of p-Terphenyl: a self-assembled supramolecular array of poly-p-phenyl nanoribbons

Luisa Barba[1], Giuseppe Chita[1], Gaetano Campi[2*], Lorenza Suber[3], Elvira Maria Bauer[3], Augusto Marcelli[5,6,7], Antonio Bianconi[2,7,8]

[1] CNR-IC Istituto di Cristallografia & Elettra-Sincrotrone Trieste, Basovizza Area Science Park 34149 Trieste, Italy

[2] CNR-IC Istituto di Cristallografia, *Via* Salaria km. 29, 00015 Roma, Italy; gaetano.campi@ic.cnr.it

[3] CNR-ISM Istituto di Struttura della Materia, *Via* Salaria km. 29, 00015 Roma, Italy

[5] CNR-ISM, Istituto Struttura della Materia, LD2 Unit, Basovizza Area Science Park, 34149 Trieste, Italy

[6] INFN, Istituto Nazionale di Fisica Nucleare - Laboratori Nazionali di Frascati, 00044, Frascati, Italy

[7] RICMASS, Rome International Center for Materials Science Superstripes, Via dei Sabelli 119A, 00185 Rome, Italy; antonio.bianconi@ricmass.eu

[8] National Research Nuclear University Mephi, Kashirskoe shosse 31, 115409, Moscow, Russia.

* Corresponding author: gaetano.campi@ic.cnr.it

## Abstract

The recent discovery of superconductivity in a metallic aromatic hydrocarbon, alkali-doped p-Terphenyl, has attracted considerable interest. The critical temperature $T_c$ ranges from few to 123 K, the record for organic superconductors, due to uncontrolled competition of multiple phases and dopants concentrations. In the proposed mechanism of Fano resonance in a superlattice of quantum wires with coexisting polarons and Fermi particles, the lattice properties play a key role. Here we report a study of the temperature evolution of the parent compound p-Terphenyl crystal structure proposed to be made of a self-assembled supramolecular network of nanoscale nanoribbons. Using temperature dependent synchrotron radiation x-ray diffraction we report the anisotropic thermal expansion in the **ab** plane, which supports the presence of a nanoscale network of one-dimensional nanoribbons running in the **b**-axis direction in the P21/a structure. Below the enantiotropic phase transition at 193 K the order parameter of the C-1 structure follows a power law behaviour with the critical exponent $\alpha = 0.34 \pm 0.02$ and the thermal expansion of the **a**-axis and **b**-axis show major changes supporting the formation of a two-dimensional bonds network. The large temperature range of the orientation fluctuations in a double well potential of the central phenyl has been determined.

## 1. Introduction

Using different procedures of alkali doping it has been recently found that p-Terphenyl $K_xC_{18}H_{14}$ shows a transition from an insulating transparent phase to an inhomogeneous metallic black phase which shows a superconductive behavior with critical temperature $T_c$ ranging from few degrees up to 123 Kelvin in different samples [1-7]. It has been proposed that the amplification of critical temperature reaching a record for metallic aromatic hydrocarbons is driven by the shape resonance belonging to the class for Fano Feshbach resonances in superconducting gaps [8]. It has been proposed that $K_xC_{18}H_{14}$ is a practical realization of a superlattice of quantum wires as described in several patents [9,10] where the critical temperature shows narrow dome if the chemical potential is fine tuned in the proximity of a topological electronic Lifshitz transition [11] in the conduction bands. The band structure calculations for $K_xC_{18}H_{14}$ [12] show the presence of the predicted Lifshitz transition [8], for x close to 3, where both a first 1D conduction band with a low effective mass, $m_1^*$, and a second conduction flat band with large effective mass $m_2^*$ cross the Fermi level. This is determined by the lattice structure of p-Terphenyl which has been described [8] as an array made of p-Terphenyl nanoribbons with large dispersion in the **b**-axis direction and low dispersion in transversal directions along the **a**-axis and **c**-axis.

In the continuously growing field of quantum materials, high interest is addressed to the manipulation of the ultrastructure of p-Terphenyl units for design of novel quantum organic materials where high temperature superconductivity could emerge [13]. According with references [14,15] the quantum coherence can resist to decoherent *attacks* of high temperatures by tuning a quantum material at a shape resonance among superconducting condensates (which belong to the class of Fano resonances or Feshbach resonances) [8,11]. In order to drive a quantum material at a shape resonance the nanoscale units should have the size close to the wavelength of electrons at the Fermi level. In this regime the chemical potential should be tuned at a Lifshitz transition for opening a neck by changing the charge density, dopants self-organization, pressure or lattice strain. Therefore the quantitative control of the variables controlling the superconducting critical temperature is needed: *i*) the charge density controlled by doping [16-18] moving the chemical potential in the conduction band [9,11]; *ii*) the lattice strain of p-Terphenyl [19,20] as in cuprates [21,22], diborides [23-25] and iron based superconductors [26]. The actual search is going on for similar doped organic compounds with the hope to reach at least the record of 203 K for the highest superconducting critical temperature, obtained so far in pressurized $H_2S$ [27,28].

Derivatives of p-Terphenyl are aromatic biological molecules present in edible mushrooms and have pharmaceutical interest for their use as immunosuppressive, anti-inflammatory and anti-tumor agents and sunscreen lotions. Moreover polyphenils have applications as laser dye in photon detectors and in light-emitting devices.

The structure of undoped parent compound (crystalline p-Terphenyl) has been object of a long standing interest [29-36]. The p-Terphenyl molecule is made of three phenyl rings connected by single C-C bond in para position. The room temperature p-Terphenyl structure with P21/a space group has been described as a packing of nanoribbons [8], made of p-Terphenyl molecules, running in the crystallographic **b** direction, as shown in Figure 1 where gray dots are carbon atoms and red dots are hydrogen atoms. Therefore the P21/a structure [8] can be classified as a realization of nanoscale architecture described in the claims of the patent for material design of high $T_c$ superconductors [9,10] made of a self-organized array of polymer quantum wires. The emergence of high temperature superconductivity in supramolecular networks of organic nanowires made of quantum wires requires the control of dopants self-organization [16-18], lattice strain [19-26] and nanoscale phase separation in the proximity of a Lifshitz transition [37-40] which need to be kept under control when the chemical potential is tuned at a Lifshitz transition [11] in the electronic structure of the conduction band [12,41,42]. The interesting scenario of high temperature superconductors tuned at a Fano resonance is that it can be obtained using multiple approaches [43-48]. The network of superconducting polymer nanoribbons is predicted to show in the metallic phase the coexistence of two electronic components, the first one in Fermi arcs with high Fermi energy $E_{F1} \gg \omega_0$, where $\omega_0$ is the pairing electron-electron attraction, and the second polaronic component in the second small electron pockets with $E_{F2} \approx \omega_0$.

In order to develop the material design of superconducting organic systems [9-10] based on doped p-Terphenyl nanoribbons as structural units it is mandatory to collect detailed information on the lattice structure of p-Terphenyl which can be obtained only with synchrotron radiation experimental methods. Here we use synchrotron X-ray diffraction to monitor the temperature range of the lattice fluctuations of the crystal structure around the order-disorder enantiotropic phase transition and to get the different anisotropic lattice thermal expansion above and below the phase transition temperature at around 193 K. We have found a highly anisotropic thermal expansion of the **ab** plane in the P21/a structure, which supports the presence nanoscale p-Terphenyl one-dimensional nanoribbons running in the **b**-axis direction .We have also determined the temperature range with large orientation fluctuations of the central phenyl ring in the proximity of the enantiotropic phase transition. The order parameter of the low temperature C-1 structure follows a power law behaviour with the critical exponent $\alpha = 0.34$. In this structure the orientation fluctuations of the central phenyl are frozen and the anisotropy of thermal expansion in the **ab** plane becomes much smaller.

2. Materials and methods

The p-Terphenyl samples, with linear formula $C_6H_5C_6H_4C_6H_5$ and molecular weight 230.30, have been synthesized in the CNR-ISM laboratories in Rome starting from p-Terphenyl powder produced by Sigma-Aldrich T3203 ≥99.5% (HPLC). The powder was dehydrated by several argon/vacuum washings at 100°C and then sealed in a Pyrex ampoule under vacuum. The ampoule was transferred in an oven and the dehydrated powder was heated up to 269°C. The temperature, controlled by means of a thermocouple positioned near to the Pyrex ampoule, was raised slowly in 5 hours and the system was maintained at 269°C for 48 hours. After completion, the ampoule was gradually cooled until reaching room temperature over a span of 7 hours.

Synchrotron diffraction experiments have been carried out in the ELETTRA synchrotron radiation facility in Trieste (Italy), at the beam line XRD1 [49]. The light was monochromatized at the wavelength of 1 Å (12.3984 KeV) by means of a double-crystal monochromator in non dispersive configuration with two Si(111) crystals, focused on the sample and collimated by a double set of slits giving a spot size of 0.2 x 0.2 mm$^2$. The sample was lodged into a glass capillary; its temperature was controlled by means of a 700 series cryocooler (Oxford Cryosystems, Oxford, UK) with an accuracy of 1 K. The thermal ramp went from 300 K to 120 K in steps of 2 K, at 120 K/hour. At every step, we hold the temperature and waited 60 s for the sample to thermalize, then collected four diffraction images rotating it under the beam of 360° in 60 s. Images were collected with a 2M Pilatus silicon pixel X-ray detector (DECTRIS Ltd., Baden, Switzerland) at a distance of 230 mm from the sample, giving the pattern a maximum resolution of 2 Å. Data were calibrated by means of a LaB6 standard, images taken at the same temperature were averaged and then integrated using the software FIT2D [50] obtaining a series of two-dimensional pattern (Intensity vs 2θ).

## 3. Results and discussion

The x-ray diffraction patterns were recorded on the x-ray diffraction beam-line XRD1 at the Elettra synchrotron radiation facility in Trieste in the temperature range between 300 and 120 K. The XRD diffraction pattern shows the monoclinic P21/a space group symmetry above the critical temperature $T_c$=193 K and the low temperature triclinic C-1 space group below $T_c$.

The lattice parameters of the room temperature P21/a space group, and of the low temperature triclinic C-1 space group are in agreement with the literature. The high temperature phase has lattice parameters a = 8.106(4) Å, b = 5.613(2) Å, c = 13.613(6) Å and β = 92.04(4)° [29-33]. The P21/a structure of p-Terphenyl shown in Figure 1 has been proposed in ref. 8 to be formed by an array of nanoscale ribbons made of p-Terphenyl molecules connected by the central phenyl groups along the **b**-axis direction.

We have performed temperature dependent X ray synchrotron diffraction measurements, cooling the sample from 300K temperature down to 120K. We have studied the anisotropic thermal expansion related to the crystallographic axis **a**, **b** and **c** in the range 120-300 K.

The 002 diffraction profile, taken as representative of all [00l] diffraction profiles is shown in the top panel of Figure 2a. The variation of the Full Width at Half Maximum FWHM is negligible and it remains very narrow (0.07 r.l.u.) at the resolution limited value at all temperatures, giving a correlation length of about 19 nm. The **c**-axis temperature variation shows that the separation between the nano-ribbons in the **c**-direction is nearly temperature independent.  The value of **c**-axis does not show major anomalies at the structural phase transition at 192.8 K. On the contrary major temperature effects are observed for the *a*-axis and **b**-axis.

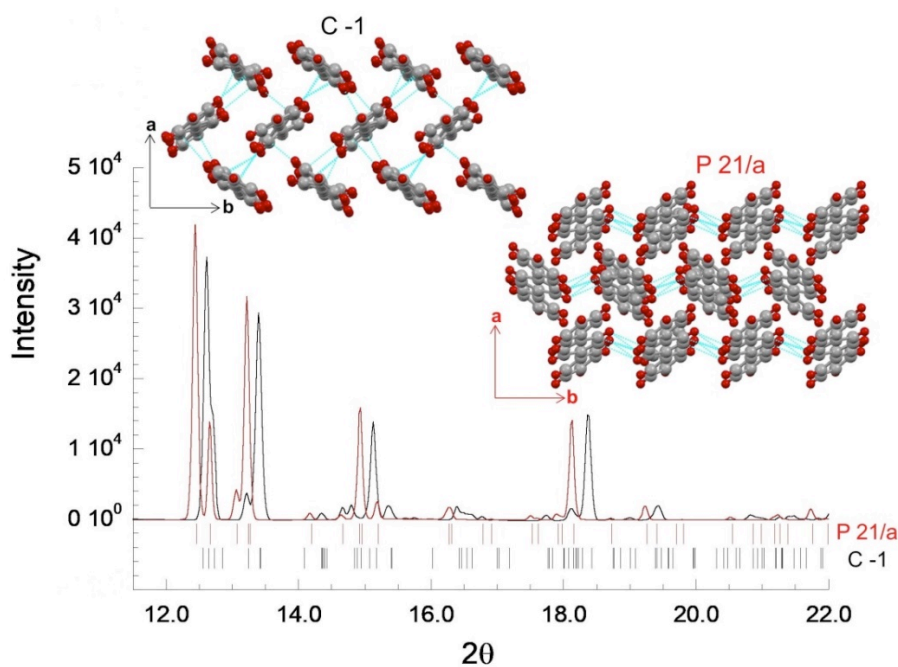

**Figure 1.** XRD patters measured at (red lines) T=300K and (black lines) T=100K. The XRD profiles have been indexed using the crystallographic structures with (CCDC reference No. 847168) C-1 and (CCDC reference No. 847173) P21/a symmetry, corresponding with the red and black markers. The p-Terphenyl structure in the C-1 space group, at T=100K and in the P21/a space group at T=300K with short bonds are illustrated. We observe the unidirectional character, along the **b**-axis, of the bonds in the monoclinic P21/a symmetry; the bonds develop in the 2D **ab** plane in the low temperature triclinic C-1 space group, with a zig-zag pathway.

The **a**-axis shows the largest contraction driven by cooling, as shown by the large temperature shift of the diffraction profiles [400] in the bottom panel of Figure 2a. The FWHM (≈0.085 r.l.u.) gives a correlation length of about 9 nm. Approaching the structural phase transition temperature it shows a negative thermal expansion below 210 K.

The [0k0] diffraction profiles probe lattice fluctuations in the direction of the nano-ribbons (see Figure 1). The color plot of the representative (040) diffraction profile of the low temperature C-1 space group, which above 193 K becomes the [020] diffraction profile of the high temperature P21/a space group, is shown in the central panel of Figure 2a. Decreasing the temperature down to the enantiotropic order-disorder transition due to the freezing of the oscillations of the central phenyl ring we observe a rapid compression of the **b**-axis anti-correlated with the **a**-axis expansion such that √(ab) does not show temperature anomalies.

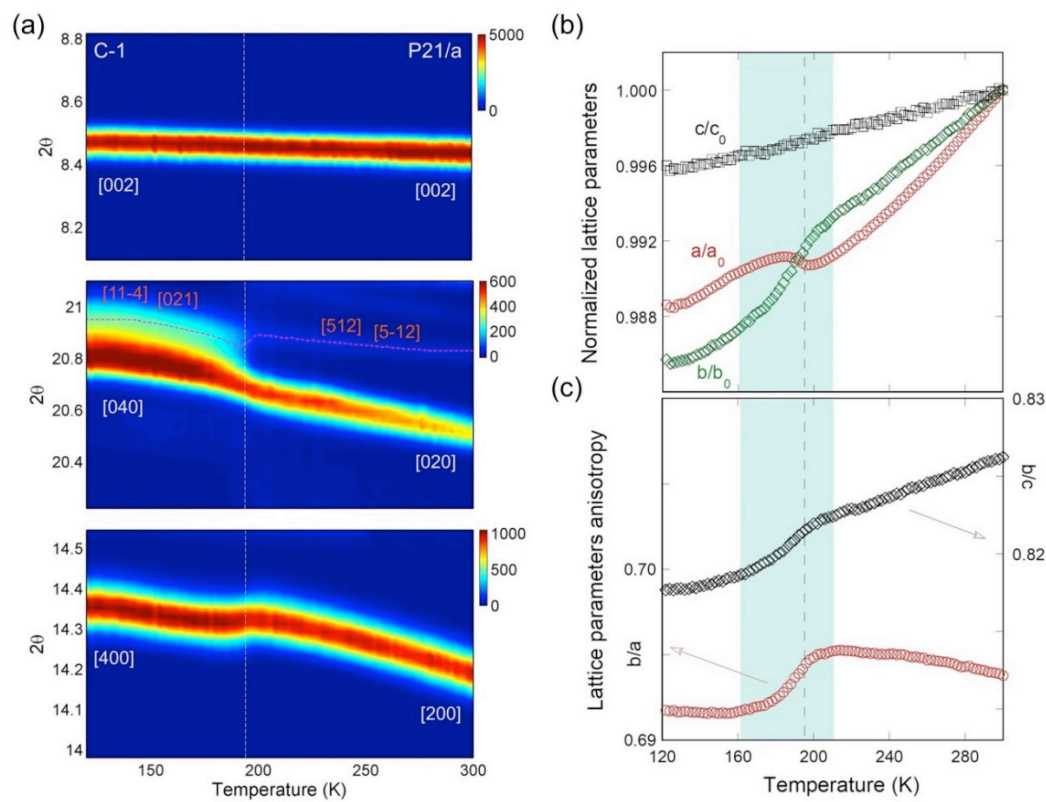

**Figure 2** (a) Color map of the x ray diffraction intensity in the 2θ reciprocal space, of the reflections classified according with P21/a symmetry notation above $T_c$ [002], [020], [200] and the according with the C-1 symmetry notation structure [002], [040] and [400] below $T_c$. Moreover we show the color plot of the [11-4], [021] reflections in C-1 notation and [512], [5-12] reflections in the P21/a symmetry as a function of temperature. (b) Lattice parameters a, b, c normalized at their value $a_0$, $b_0$, $c_0$, at 300K. The anisotropic cooling compression of the crystal is clearly indicated by the temperature variations of the b/a and b/c axis ratio shown in panel (c). We can see the different anisotropic thermal expansion above and below the critical temperature 192.8±0.5 K. The width of the transition spans a temperature range going from 135 K to 250 K.

The central phenyl unit of p-Terphenyl has twisting degrees of freedom in the P21/a structure which results in a dynamical instability for its rotation in a double well potential. Therefore it is possible that charge carriers in doped p-Terphenyl could couple to lattice fluctuations in a double well potential and charge density waves could appear as in cuprates and iron based superconductors.

Decreasing the temperature down to the order-disorder enantiotropic phase transition below 193 Kelvin the structure is described by the C-1 symmetry as observed in x-ray and neutron diffraction experiments. In the low temperature C-1 symmetry the fluctuations of the central phenyl are frozen with ordering of different local structural polymorphic conformations in different crystal sites breaking the nano-ribbons 1D linear chains.

The room temperature thermal expansion in the **c**-axis direction has been obtained by a linear fit of the temperature variation of the **c**-axis from the data in the range 250 K to 300 K with the linear curve $[c_0 - c(T)]/c_0 = C_c(300 - T)$ where $c_0$ is the *c*-axis value at 300K. We have found a very small **c**-axis thermal expansion coefficient $C_c = 2.1 \pm 0.5 \; 10^{-5}$ K$^{-1}$. The latter expansion coefficient is nearly temperature independent down to 120 K showing that the system is very stiff in the **c**-axis direction and no anomalies are observed at the enantiotropic phase transition. On the contrary the temperature dependence of the **a**-axis and **b**-axis show large anti-correlated variations around the enantiotropic phase transition.

Linear thermal expansions of the **a**-axis and **b**-axis near room temperature can be approximated

$$[a_0 - a(T)]/a_0 = C_a(300 - T); \; [b_0 - b(T)]/b_0 = C_b(300 - T)$$

where $a_0$ and $b_0$ are the **a**-axis and **b**-axis values at 300K. The thermal expansion coefficients are found to be $C_a = 10 \pm 0.3 \; 10^{-5}$ K$^{-1}$ and $C_b = 7.5 \pm 0.3 \; 10^{-5}$ K$^{-1}$ showing the large lattice anisotropy in the **ab** plane. This anisotropy results from the difference in bond strengths. The intermolecular p-Terphenyl bonds in the **b**-axis directions are much stronger than the bonds that connect p-Terphenyls along the **a**-axis in the high temperature P21/a structure. This results support the model of p-Terphenyl nanoribbons in the in the P21/a space group with stronger bonds between the central phenyl rings which run in the **b**-axis direction as shown in Figure 1.

In the low temperature C-1 phase the thermal expansion coefficient in the **a**-axis direction decreases and at around 150 K it becomes similar to the thermal expansion in the **b**-axis direction $C_a = C_b = 5.8 \pm 0.5 \; 10^{-5}$ K$^{-1}$. The anti-correlated thermal behavior of the **a**-axis and **b**-axis around the enantiotropic transition from P21/a to C-1 structure occurs in a wide temperature range between 240K and 160K, indicated by the shaded area in Figure. 2.

In Figure 3 we focus on the powder pattern around the [0k0] reflections, more precisely on the [040] and [020] peaks in the C-1 and P21/a symmetry, respectively. Figure 3a shows a 3D plot of the XRD pattern around 2θ=20.7° of the [020] and [021], [11-4] reflections in the P21/a evolving in the [040] and [5-12], [512] reflection in the C-1 space group, as a function of temperature. At the disorder-to-order transition we observe an strong increase of both and amplitude and FWHM of [0k0] peaks, as shown in Figure 3b and 3c, respectively.

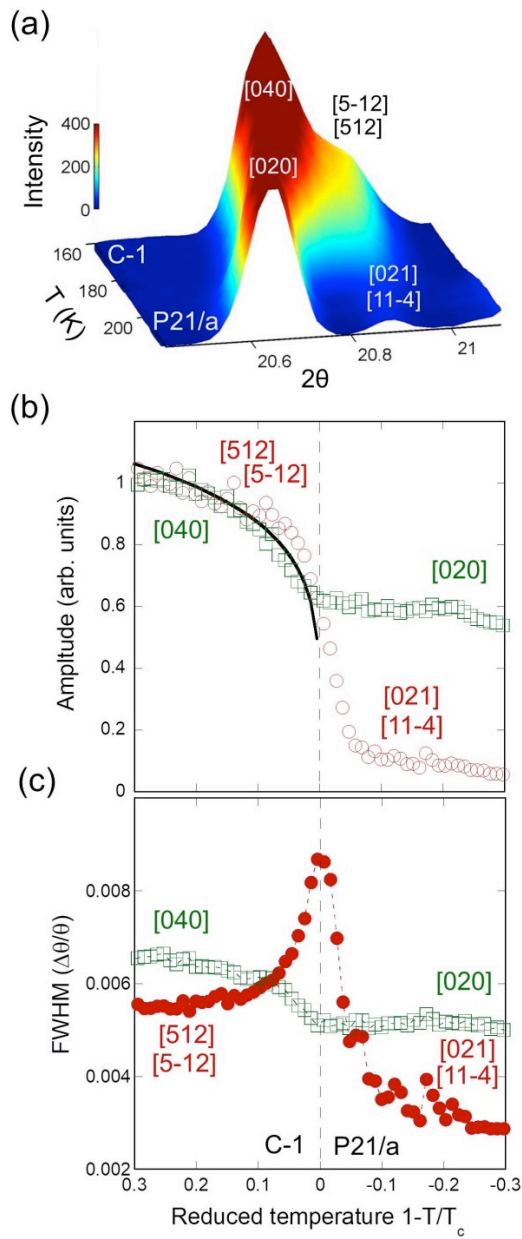

**Figure 3 (a)** Color plot around the structural phase transition of the x-ray reflection profiles [040] and [5-12], [512] in the C-1 low temperature structure evolving at high temperature to the [020] and [11-4], [021] peaks in the high temperature P21/a structure showing large structural fluctuations of the two competing phases over a wide temperature range of 115 K centered around the critical temperature Tc. **(b)** Amplitude and **(c)** FWHM ($\Delta\theta/\theta$) obtained by fitting the profiles of diffraction peaks both in the C-1 low temperature structure and in the high temperature P21/a structure. In (b) we show the best fitted curve of the diffraction intensity $I_0 (1-T/T_c)^\alpha$ with $\alpha=0.34\pm0.02$ and critical temperature $T_c=192.8\pm0.5$ K.

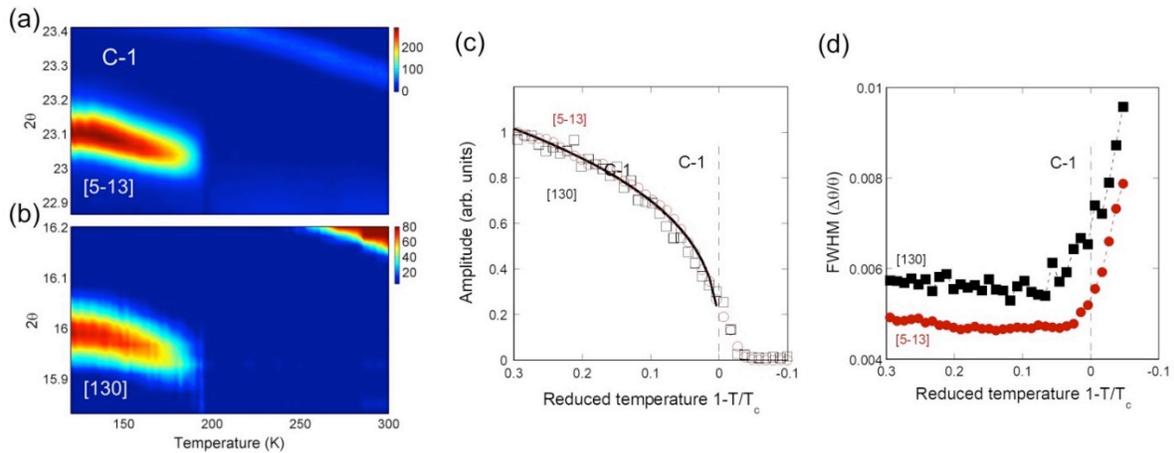

**Figure 4**. Color plot of the (a) [5 -1 3] and (b) [1 3 0] reflections in the C-1 symmetry at low temperature of the disappearing in the high temperature P21/a structure. (c) Amplitude and (d) FWHM obtained by peak fitting of [130] and [5-13] peak profiles in (b). We show the best fit following the power law behavior at the phase transitions with $(1-T/T_c)^\alpha$ with α=0.34±0.02 and $T_c$=192.8±0.5 K.

At the same time the FWHM of the convolution peaks of [021], [11-4] triclinic reflections and of [5-12], [512] monoclinic reflections diverges at $T_c$, as expected at a phase transition. This type of transition is well described by the power law decay of the peak amplitude with lowering temperature, reaching zero at $T_c$

$$I(T) = I_0(1 - T/T_c)^\alpha \qquad\qquad 1$$

as shown in Figure 3b. In this figure the black line represents the best fitted curve following Eq. 1 behavior, giving $T_c$ = 192.8±0.5 K and α = 0.34±0.02 .

The enantiotropic order-disorder transition is directly probed by the appearing of new superstructure peaks below $T_c$ = 192.8±0.5 K as shown in Figure 4. They are represented by the color temperature plot of the [5-13] and [130] diffraction profiles (Figure 4a and 4b, respectively) appearing only in the C-1 symmetry at low temperature. The structural phase transition from the high-temperature P21/a phase to the low-temperature C-1 phase is clearly observed in Figure 4c and 3d where we show the evolution of the amplitude and the FWHM of the [5-13] and [130] diffraction peaks, as a function of 1-T/$T_c$ where $T_c$ is the phase transition temperature. Also in this case we find a transition well described by the power law $I_0(1-T/T_c)^\alpha$ with $I_0$ constant, critical exponent and α=0.34 and $T_c$=192.8±0.5 K. At this temperature, as expected, we find the divergence of the width peaks (Figure 4d).

In the C-1 space group symmetry we associate the lack of thermal expansion anisotropy with the disappearing of the nanoribbons in the **b**-axis direction and the formation of similar arrays of bonds in the **a**-axis and **b**-axis direction in the **ab** plane.

### 4. Conclusions

We present the temperature evolution of p-Terphenyl structure by means of synchrotron X ray diffraction. We collected XRD patterns during a cooling ramp from 300K down to 120K. Monitoring the lattice parameters behavior we got evidences of an anomalous thermal expansion which is quasi uniaxial at high temperatures, above $T_c$, while becoming biaxial at low temperature, below $T_c$. This transition takes place in a wide temperature range following a power law behavior below $T_c$. The understanding of structural elastic properties of p-Terphenyl is one of the most important issue to control the doped metallic p-Terphenyl and to design new supramolecular organic superconducting systems. In fact the control of supramolecular self-assembly of a given molecular building block provides a road map for material design of intricate structures containing 3D networks made of 1D organic quantum wires, opening new venues for modern nanotechnology. Soft and biological matter provide a wide landscape of multiscale supramolecular self-assembly of quantum nano-scale modules therefore it represents a promising roadmap to get room temperature superconductors. Quantum coherence at high temperature shown in high temperature superconductivity has found to emerge in different supramolecular arrays of metallic quantum wells, like in diborides [25] and iron based superconductors [26], or supramolecular arrays of metallic quantum wires like in cuprates [14] and pressurized sulfur hydride [22].

The simplest case of high temperature superconductivity driven by shape resonance is provided by $MgB_2$ [25] which is a supramolecular self-assembly of an array of quantum wells, formed by stacks of 2D graphene-like atomic boron layers. Also iron based superconductors [26] are a formed by self assembly of FeX (X=As, Se) quantum wells. However the highest critical temperatures occur in self-assembly of supramolecular array of quantum wires like in cuprates [14,15] and pressurized sulfur hydride [22] tuning the chemical potential at a Lifshitz transition [11,51,52]. Therefore advances in the materials science control of the self-organization of arrays of metallic poly-phenyl quantum wires is needed to design high temperature organic superconductors.

It was known that the central ring of each p-Terphenyl molecule lies at an angle to a plane defined by the outer rings just because of the competition among inter-ring H–H bonds with ortho-hydrogens which force the adjacent rings to be non-coplanar, and π orbitals on the phenyl rings which prefer a coplanar arrangements. In the **ab** plane there are two possible conformations: one with the central ring of the p-Terphenyl molecule twisted in one sense, and the other twisted oppositely. This order doubles the C-1 cell along a and b respect to the high temperature P21/a cell. Above $T_c$, our results also support the formation of 1D nanoribbons so that the average structure unit cell is the P21/a cell, but the two senses of twisting give a dynamical fluctuation in a double well potential. In the low temperature triclinic C-1 space group the p-Terphenyl nano-ribbons are broken by freezing the oscillations of the central phenyl group at  different tilting angles. This makes the studied transition from the high temperature symmetry P21/a to the low temperature C-1 symmetry quite complex.

Here we have shown that networks of one dimensional (1D) nanoribbons in p-Terphenyl in the P21/a structure are characterized by an unusually large anisotropy of the thermal expansion in the **ab** plane in agreement with anisotropic pressure compressibility [20] clarifying  one of the key feature of self-organization of molecular polymorphism and supramolecular isomerism in p-Terphenyl. The thermal anisotropy in the plane decreases and it is reversed in the low temperature C-1 phase indicating the formation of a 2D network of bonds in the **ab** plane associated with cooling induced compressive strain. Work is in progress to measure the temperature dependent supramolecular self-assembled superstructures of alkali doped organic of polyphenyls determined by potassium dopant interactions with p-Terphenyl and intermolecular C-H $\pi$ bonds.